\documentstyle[preprint,aps,eqsecnum]{revtex}
\tighten
\begin{document}
\title{\Large{\bf The String Equation and Solitons}}
\author{{\bf  S. P. Novikov}}
\address
{Landau Institute for Theoretical Physics, Moscow, RUSSIA}
\maketitle

\vskip 80pt

\begin{center}

Based on lectures delivered at the ERICE CHALONGE School, `STRING GRAVITY AND

PHYSICS AT THE PLANCK  SCALE', 8-19 September 1995,

to appear in the Proceedings edited by  N. S\'anchez, Kluwer Publ. Co.

\end{center}
\newpage
\vspace{0.5cm}
\vspace{0.5cm}
\centerline{\bf Lecture 1: Periodic 1- and 2-dimensional Schrodinger Operators}

\centerline{\bf Riemann surfaces, Nonlinear Equations}

\vspace{0.5cm}
{\bf Introduction}. We are going to present here some brief survey of
the results of Theory of Solitons (see  \cite{N,AS,DKN}) from the
viewpoint of periodic theory including some new results in the theory
of 2-dimensional periodic Schrodinger Operators.

A remarkable connection of some very special but highly nontrivial
 nonlinear (especially one-dimensional)
systems with spectral properties of one-dimensional linear Schrodinger
Operators was discovered in 1965--68
in the series of works \cite{KZ,GGKM,L} for the
famous KdV equation
$$u_t=6uu_x-u_{xxx}\; .$$

This connection is based on the identification
of KdV with Heisenberg type equation for the linear operators
(''Lax representation''):
$$L_t=[A,L], L=-\partial_x^2+u(x,t), A=-4\partial^3_x+3(2u\partial_x+u_x)$$

This equation generates an effective GGKM integration procedure
(''Inverse Scattering Transform'') for KdV
in the class of rapidly decreasing functions $u(x)\rightarrow 0,
|x|\rightarrow\infty$, using the solution of inverse scattering problem
for the one-dimensional Schrodinger Operator. A lot of important results
were extracted from
this method in the 70-s, using
traditional methods in the modern Theoretical and Mathematical Physics like
exact multisoliton solutions, asymptotics
for $t\rightarrow+\infty$, action-angle type variables for the
rapidly decreasing KdV, new classes of systems integrable by the
same trick (see \cite{AS,N}). Many groups participated in this development,
including  Zakharov, Shabat, Lamb, Faddeev, Ablowitz, Kaup, Newell, Segur,
        Henon, Flashka, Manakov, Moser, Calogero and others. They found a lot
of new important    ODE-s and PDE-s
(0+1,1+1 and even 2+1)-- systems with Lax-type representation, including
some special cases of Einstein equations and Self-Duality equation for
the Yang-Mills fields (the last system is 4-dimensional).

{\bf Periodic Solitons and Algebraic Geometry}.

We discuss here a different part of  development of this theory,
based on the ''Triangle'' with  following vertices:

1.Algebraic Geometry, associated with Riemann surfaces
and their
$\Theta$-functions (which never has been used in Applied Mathematics
and Physics before);

2.Spectral theory of  Schrodinger operator
on the real line $x$ with periodic (quasiperiodic) potential;

3.Periodic problems for KdV and higher analogues.

This development started from the work \cite{n1} in 1974 and was
realized completely (in the case of KdV) by the present author, Dubrovin,
Its, Matveev, Lax, Mckean, van Moerbeke in 1974-75 (see  full
 survey of this theory in \cite{DMN,N,DKN}).

In the periodic case (I remind that the potential $u$ is periodic here,
not the eigenfunction) the corresponding inverse spectral problem
for Schrodinger operators never has been solved before the KdV theory.
It was solved only in 1974-5 as a part of KdV theory, using
the deep connection of KdV and Schrodinger operator. This solution was based
on the new ideology, considering KdV and its higher analogs
as a symmetry theory for the Schrodinger operator ( for its
spectral theory): eigenvalues of Schrodinger operators are the integrals
of motion for all higher KdV systems. We have an infinite-dimensional
commutative symmetry group for any functional $F{u(x})$, if it depends
on the spectrum of the operator $L=-\partial+u$ only. Such functionals
played an important role in the definition and studying of  finite-gap
potentials (below). Completely different important example  was
found later: a very
well known Peierls Free Energy functional in the mean field approximation
for some electron-phonon systems, describing the so-called ''charge density
waves'' in some quasi-one-dimensional media. Its exact integrability was
discovered by the group of physicists in collaboration with experts in the
soliton theory: Belokolos, Dzyaloshinski, Gordyunin, Brozovski, Krichever
(see, for example, in the survey \cite{DKN}).

An important generalization of
finite-gap solutions for
some special 2+1 systems (like KP) was done by Krichever
(see \cite{kr}), who extended very far an algebraic part of
periodic theory. Many people worked in this area later. The references
may be found in the books \cite{N,DKN}.

This approach is based on the special finite-dimensional
families of exact solutions, whose $x$-dependence is
specified by the Commutativity Relation
of 2 different linear OD operators $[C,B]=0$. For 1+1 (or $x,t$)-systems
the corresponding operator $C$ is necessarily equal to the operator
$L$ in the Lax pair for our system. The operator $B=\sum_ic_iA_i$
is some linear combination of corresponding $A$-operators for the so-called
''Higher KdV''systems, associated with the same 1-dimensional
Schrodinger Operator in the case of KdV.  Commutativity Relation is
equivalent to the family of
 Completely Integrable Hamiltonian OD systems
in the variable $x$, admitting
some very useful ''Lax-type representations''
$$\Lambda(\lambda)_x=[\Lambda,Q]$$

for $2\times 2$--matrices, depending on the additional parameter
$\lambda$. Riemann surface $\Gamma$ appears as a polynomial equation
$$\det[\Lambda(\lambda)-\mu]=0$$

whose coefficients are the integrals in $x$.

{\bf Example}: In the simplest case of stationary waves for KdV
we have $C=L,B=A$,
a Riemann surface $\Gamma$ with genus 1, extracted from the matrices:
\begin{eqnarray}
\Lambda=\left(\begin{array}{cc}-u_x&2u+4\lambda
\\-4\lambda^2+2\lambda u-u_{xx}+2u^2&u_x\end{array}\right),Q=\left(
\begin{array}{cc}0&1\\u-\lambda&0\end{array}\right)
\end{eqnarray}

Generic finite-gap solutions $u(x)$ are periodic (or quasiperiodic) in $x$.
They
can be written by the formula
$$u=-[2\log\Theta(Ux+Wt+U_0)]_{xx}+Const$$

with Riemann $\Theta$-function, vectors $U,W$ and constant $C$ determined by
the Riemann surface $\Gamma$; these potentials
 have a remarkable Spectral Property: Corresponding Schrodinger
Operator $-\partial_x^2+u(x)=L$ has only
a finite number of gaps in the Spectrum on the line,
whose endpoints are exactly
the branching points of  Riemann surface above.

The Spectral Problem  of Bloch
$$L\Psi=\lambda\Psi,\Psi(x+T)=e^{ipT}\Psi(x)$$

is completely solvable in $\Theta$-functions.

This class of functions generates the ''Finite-Gap Solutions'' of KdV
\cite{DMN}.
This Family is dense in the class of all smooth periodic functions
 --see\cite{MO}.

For the 2+1-dimensional KP--system
the corresponding Lax-Zakharov-Shabat operators are
$$L=\sigma\partial_y-\partial_x^2+u(x,y,t),\sigma^2=\pm 1,
 A=-4\partial_x^3+3(2u\partial_x+
u_x)+w$$

There is no relation here between the form of  linear
OD operators $[C,B]=0$, describing  special solutions,
associated with Riemann surfaces, and  Lax operators $L,A$. All family of
''Krichever Solutions'' for KP
is much more broad than  finite--gap families in the case of 1+1 systems,
They have more or less the same analytical form as above,
but the class of parameters (compact Riemann
surfaces with marked point) is unrestricted.
This class of solutions was used later several
times for the different goals: for the solution of the classical problems of
the theory of $\Theta$--functions
(like new approach to the Riemann--Schottki Problem, started in  \cite{D1}
and finished by Shiota),
for some applications in the Conformal 2-d Field Theory  and in the theory
of bosonic strings on the base of
new beautiful algebraic \cite{SMJK} and functional \cite{KN2}
interpretations. An extension of this class,
associated with such modern aspects of Algebraic Geometry as
holomorphic vector bundles over algebraic curves and their
deformations, was constructed in \cite{KN}.

{\bf Solitons and strings}:
In particular, in the works \cite{KN2} we realized the following program.
As everybody know, in the late 60-ies and early 70-ies a large group of
physicists (Veneziano, Virasoro, Alessandrini, Mandelstam and many
others) developed the very beautiful theory of the bosonic quantum strings.
They used  standard operator quantization, decomposing  fields
in the Fourier series and  replacing c-numbers by operators with
standard canonical commutators. This program was effectively
realized for the ''zero-loop'' or
''tree-like diagrams'' only (i.e. for the processes,
described by the Riemann surfaces of the zero genus). The so-called
Virasoro algebra and its representations played an important role in
these constructions. This program stopped because nobody was able
to quantize fields in such a way for the multiloop case (i.e. for the
Riemann surfaces of nonzero genus). In early 80-ies Polyakov
solved the problem of quantization of bosonic strings using a
functional (''path'') integral. No objects like Virasoro algebra
appear in this approach. In the works \cite{KN2} we constructed a
right analog of the Fourier-Laurent series on the Riemann surfaces,
using the analytical constructions of the Soliton Theory. After that,
an operator quantization of strings was done very easily. Some beautiful
analogs of the Virasoro algebra appeared here. This area was not active
in the last 3 years, so we shall not discuss it here.

In the second lecture we
shall discuss a completely different deep connection between  solitons
and strings.

Let me start now the main subject of this lecture.

\bigskip

{\bf Topologically Trivial
Periodic 2-dimensional Schrodinger Operators and Riemann surfaces.}.

\bigskip

I have no intention to discuss here all the subjects above.
 My goal is to explain  some less popular ideas, associated with
2-dimensional Schrodinger Operator, in connection with new work of
the present author and A.Veselov (in preparation).

It is more or less obvious, that there is no nontrivial Lax
equations associated with 2-dimensional Schrodinger operator
$-2L=(\partial+A_c)(\bar{\partial}+B_c)+2V$. Here $A_c,B_c$
are the components of vector-potential, $V$ is a scalar potential,
$\partial=\partial_x-i\partial_y, z=x+iy$

However, nontrivial integrable nonlinear systems can be obtained
from the different equation ({\bf $''L-A-B$--triple''}),
 which appeared and was investigated since
 1976 (see \cite{M,DKN3}). The inverse spectral problem
for double--periodic Schrodinger operators $L$ is
 associated with one energy level only $(L\Psi=0)$ in the approach.
 $L-A-B$ triple equation has a form:

$$L_t=[A,L]+BL=LA+(A+B)L$$

which implies something like Lax representation, corresponding to one
 energy level:
$$(L_t-[A,B])\Psi=0,L\Psi=0$$

This representation leads to some beautiful 2-dimensional analogs of KdV,
 containing KP as some degenerate limit:
$$u_t=(\partial^3 +\partial P)u(x,y,t)+C.C.,\bar{\partial}u=3\partial P,
u=\bar{u}$$

and corresponding analogs of ''Higher KdV'' systems. Nontrivial
exact solutions of this nonlinear systems and periodic
Schrodinger operators with zero magnetic field
$-2L=\partial\bar{\partial}+2V$ and solvable Bloch
problem $L\Psi=\epsilon_0\Psi$
for one energy level
were found by the present author in collaboration with
Veselov in 1984 (see   \cite{N3}.
Our Riemann surface $\Gamma$ in this case is exactly a Complex
Fermi Curve.  Our Bloch wave function $\Psi$
can be expressed through the so called ''Prym''
$\Theta$--functions, which are more complicated than the standard ''Jacobian''
$\Theta$--functions
in the case of KP above. Generic  complex Fermi curve has
 infinite genus.
This theory was developed by Krichever in 1989--90,
who proved that our exactly integrable
 class (with Complex Fermi Curve of finite genus)
is dense in the class of all double periodic potentials.
Rapidly decreasing class also was investigated by Grinevich,
Manakov, R.Novikov and the present author in 1987--89.
 It is interesting
that for the two-dimensional Schrodinger operator periodic inverse problem
was solved earlier than rapidly decreasing inverse scattering
problem (based on the data, associated with
one energy level). There exist simple
rational potentials (found by Grinevich in 1988),
for which the scattering amplitude is identically equal to
zero for one energy level.

\bigskip

{\bf Topologically Nontrivial Schrodinger Operators}.

\bigskip

All this class of integrable Schrodinger operators, associated with
nonlinear systems, Riemann surfaces and $\Theta$--functions,
contains only  Schrodinger operators with
''topologically trivial'' magnetic field: it has a
 ''Chern class'' (i.e.
 magnetic flux through the elementary cell in the double periodic case)
equal to zero.

Completely different class of Schrodinger operators with exactly integrable
ground level (which is highly degenerate) was found 15 years ago
in rapidly decreasing
(\cite{AC}) and periodic (\cite{DN}) cases.
It corresponds to the nonrelativistic Pauli operator for spin 1/2
in the magnetic field, orthogonal to the plane, and zero electric potential.
The ground energy level is equal to zero in this case.
In particular, for the periodic case \cite{DN}, this level is isomorphic
 to the first Landau level in the constant magnetic field
 with the same magnetic
flux through the elementary cell. If this flux is an integer, the so called
''Magnetic Bloch functions''
  were found analytically through
the elliptic functions for all this class in \cite{DN}:
$$\Psi(x,y)=e^{\phi}\sigma(z-a_1)\ldots\sigma(z-a_n)e^{az}$$
$$\Delta\phi=-H$$

Here $H$ is a a magnetic field, $a$ is expressed through the constants
$a_1,\ldots ,a_n$ and $H$.

\bigskip

{\bf Cyclic and semicyclic chains of the Laplace transformations. New results
of the present author and A.Veselov}.

\bigskip

For the unification of these two theories the present author in collaboration
 with Veselov used an idea of ''Cyclic Chains'' of Laplace transformations.
Let me point out that the theory of cyclic chains of Backlund transformations
for 1-dimensional Schrodinger operator was developed in the beautiful
work of Shabat and Veselov \cite{SV} (some first observations were
found in \cite{W}). In the early XIX century, Laplace constructed the
transformations of second order linear PDE for some goals in geometry.
I would like to point out that the ''Laplace transformation''
acts on the solutions of the equation
$L_0\Psi_0=0$ for the two-dimensional Schrodinger operator $L_0$
$$L=L_0=-1/2(\bar{\partial}+B_0)(\partial+A_0)+V_0$$
by the formula
$$\Psi_1=(\partial+A_0)\Psi_0,A_1=A_0-(\log V_0)_z, B_1=B_0,
V_1=V_0+H_1$$
Here the magnetic field $H_0$ is equal to $2H_0=B_{0z}-A_{0\bar{z}}$
and $H_1,V_1$ are the magnetic field and scalar potential for
the operator $L_1$ respectively, such that
$$L_1\Psi_1=0,H_1=H_0+1/2(\log V_0)_{z\bar{z}}$$

The requirement that the chain of Laplace transformations is periodic
leads  to the beautiful elliptic partial
differential equations for the magnetic fields
and scalar potentials of all operators in the chain.
This problem was posed and studied for the first time in the XIX
century by Darboux
(the operator $L$ and the corresponding nonlinear systems are
always hyperbolic in geometry, but formal calculations
are the same).
Some useful formal calculations were done by  Tsiseika in 1920-s.
We applied this stuff to the theory of (elliptic) 2-dimensional Schrodinger
operator.
Globally nonsingular double periodic solutions
of this system in our ''elliptic'' case
give two-dimensional Schrodinger operators with Complex Fermi
Curve of finite genus (Algebro-Geometric operators, as above, in
the case of topologically trivial magnetic field.)

Especially beautiful well-known integrable systems appear in the case
 $H_0=H_n=0$
for the periods $n=3,4$. Let $V_0=\exp f$. We have:
$$ \Delta f=-2(e^f-e^{-2f}),n=3,\Delta f=-4sh f,n=4$$

In the topologically nontrivial double periodic case, when the magnetic flux
$[H]$ is nonzero, no cyclic chain is possible. Let be $[H]>0$.
Instead of cyclic chains we consider the {\bf Semicyclic Chains}
and {\bf Quasicyclic chains},
 satisfying  one of the two following conditions:

1.Semicyclic chains
$$H_0=H_n, V_n=V_0+n[H_0]$$

It leads to some operators with special algebraic properties:

Eigenfunctions of two different energy levels 0 and $n[H_0]$
are connected by the operator
$$\Psi_n=(\partial +A_{n-1}\ldots)(\partial+A_0)\Psi_0$$

For $n=2$ this condition leads to the equation
$$\Delta f_0=a-bsh f_0, b> 0\; ,$$

which has a lot of double periodic nonsingular
solutions. However, these levels, equal to zero and to $n[H_0]$
can be out of the spectrum, so this connection is formal.

\bigskip

{\bf 2.Quasicyclic chains}

\bigskip

Another, more interesting analog of cyclic chains, we obtain from the
 condition, that
for $n=0$ and for some other value of $n$ the operators $L_0,L_n$
belong to the class \cite{AC,DN} up to the shift of energy
$$V_0=H_0, V_n=H_n+n[H_0]$$

For $n=1$ there exists only a constant solution of this equation, but
for $n=2$ it  leads to the elliptic PDE:
$$\Delta g=4([H_0]-e^g), e^g=V_0$$

This equation has a lot of real nonsingular double periodic solutions
on the plane. The corresponding Schrodinger operators $L_n$
have two highly degenerate energy levels:

1.Ground level, equal to zero, with magnetic
Bloch eigenfunctions written above (by the results of \cite{DN})

2. Second integrable level, equal to $n[H_0]$, with magnetic Bloch
eigenfunctions of the form

$$\Psi_n=(\partial+A_{n-1})\ldots(\partial+A_0)\Psi_0$$

Here $\Psi_0$ is a magnetic Bloch eigenfunction with zero energy
level for the operator $L_0$, written by the same formula, but with
different magnetic field:
$$H_0=e^{f_0},H_2=2[H_0]-e^{f_0},n=2,[H_0]>0$$

These levels are isomorphic to the
''Landau levels'' with numbers 0 and $n$ of the
operator $L_0$ with constant
homogeneous magnetic field.

In both levels the magnetic Bloch functions can be calculated through the
elliptic functions, written above. H. de Vega pointed out that
our nonlinear equation for $n=2$ appeared already in the 70-s
as an ''Instanton Reduction'' for the Landau-Ginzburg equation
for the critical value of parameter [separating between  superconductors
of the first and second kind (see \cite{DeV})] .

\pagebreak

\vspace{1cm}
\centerline{\bf Lecture 2: Theory of Solitons and String Equation}
\vspace{0.5cm}
In the modern  terminology, ''String Equation'' means exactly the equation
$$[L,A]=1$$
for some linear OD operators (people call it also a ''Heisenberg Relation'').

This strange terminology appeared
in 1989-90 years after the well-known works of Gross,
Migdal, Brezin, Kazakov, Douglas, Shanker, David and many
others (see \cite{GMi,BK,DS,D}).

A partition function and Free energy of
$N\times N$--Matrix Models in Statistical Mechanics in some very special
''Double-Scaling Limit'', when the size of matrices is going to infinity
$N\rightarrow\infty$, have probably a beautiful interpretation in
the String theory, which was conjectured by the above-mentioned physicists.
These Matrix Models and their ''string limit''
have a deep connection with the Theory of Solitons, which was the most
interesting mathematical discovery of these authors. An analysis
of this limiting process was done by Its and others (see \cite{I}).
Famous KdV type
Systems of the Soliton Theory play here a role of ''Renormalization Group'',
like in Quantum Field Theory. However, completely different
classes of special solutions are needed here. It is good to point out,
that the Integrability in the sense of Lax--type representation leads to
the effective results only for 2 classes: periodic (quasiperiodic) and
rapidly decreasing. Sometimes in the theory of Solitons people needed
in the self-similar solutions for asymptotic methods and so on. A
beautiful idea
to study them was invented by Flashka and Newell about 1979 (it was developed
 by the Japanese school and later used by the Leningrad and Ufa schools
for asymptotical studyings)--see\cite{FN,Jap,I2,Ko,Ki}. However,
this approach is
complicated; it gives a very few number of the effective results. We have here
exactly this case.
In particularly, the computation of  Free energy
can be reduced to the Painleve'--1 equation in the simplest nontrivial case
$$u_{xx}-3u^2=x$$

In fact, it is equal in this limit to the special real ''Physical solution''
on the positive halfline $x\leq 0$ with asymptotics
$$u(x)\sim +\sqrt{(-x)/3}$$

This nonlinear equation is equivalent to the algebraic ''String Equation''
or ''Heisenberg Relation'' above $ [L,A]= 1 $ for the same OD operators
which give a Lax pair for the ordinary KdV equation (see Lecture 1).
This observation leads to some analog of the  Lax representation for this
Painleve'-1 equation. Several new approaches were developed for
the investigation
of this equation on the base of technics of the Theory of Solitons
(see \cite{n2,Mo,Krich,IKF,GN2}). I  presented
in this lecture the ideas of the last
joint work of myself with Grinevich (\cite{GN2}),
where a special isomonodromic method for
the studying of the physical solution was developed.

\pagebreak


\begin{thebibliography}{99}

\bibitem{N}Novikov S.P., Manakov S.V., Pitaevski L.P., Zakharov V.E.
{\em Theory of Solitons.Plenum Press, 1984}

\bibitem{AS}Ablowitz M., Segur H.
{\em  SIAM, Philadelphia, 1981}

\bibitem{DKN}Dubrovin B.A., Krichever I.M., Novikov S.P.
{\em Integrable Systems. Encyclopedia Math Sciences, Dynamical Systems,
vol 4 (Edited by V.Arnold and S.Novikov), Springer.}

\bibitem{KZ}Kruskal M., Zabusky N.
{\em  Phys Rev Lett., 1965, vol 15 pp 240--243}

\bibitem{GGKM}Gardner C., Green J., Kruskal M., Miura R.
{\em   Phys Rev Lett., 1967, vol 19 pp1095--1097}

\bibitem{L}Lax P.
{\em  Comm Pure Appl Math, 1968, vol 21 iss 5 pp 141--188}

\bibitem{n1}Novikov S.P.
{\em Functional Analysis Appl., 1974, vol 8 iss 3 pp 54--66  }

\bibitem{DMN}Dubrovin B.A., Matveev V.B., Novikov S.P.
{\em Russian Math Surveys, 1976, vol  31 iss 2 pp 55--136 }

\bibitem{MO}Marchenko V.A.
{\em  Kiev, Naukova Dumka, 1977}


\bibitem{kr}Krichever I.M.
{\em Soviet Math Doklady, 1976, vol 227 iss 2 pp 291--294;
Russian Math Surveys,
1977, vol 32 iss 6  pp 180--208 }

\bibitem{D1}Dubrovin B.A.
{\em Russian Math Surveys, 1981, vol 36 iss 2 pp  11--80 }

\bibitem{SMJK}Sato M., Miwa T., Jimbo M.
{\em   Publ. RIMS: 1978,vol 14 p 223; 1979, vol 15 pp 201, 577, 871;
1980, vol 16 p 531}

\bibitem{KN2}Krichever I.M., Novikov S.P.
{\em Functional Analysis Appl., 1987 vol 21 iss 2 pp  46--63 ;
1987, vol 21 iss 4 pp  47--61;
1989, vol 23 iss 1 pp  24--40;
in the collection of papers ''Physics and Mathematics
of Strings'', dedicated to the memory of V.Kniznik
(edited by Friedan, Brink and Polyakov), Singapour}

\bibitem{KN}Krichever I.M., Novikov S.P.
{\em  Russian Math Surveys, 1980, vol  35 iss 6 pp 47--68  }

\bibitem{M}Manakov S.V.
{\em Russian Math Surveys--Notes of the Moscow Math Society, 1976, vol 31
iss  5 pp 245--246 }


\bibitem{DKN3}Dubrovin B.A., Krichever I.M., Novikov S.P.
{\em Soviet Math Doklady, 1976, vol  229 iss 1 pp 15--18  }

\bibitem{N3}Novikov S.P., Veselov A.P.
{\em Soviet Math Doklady, 1984 vol 279 iss 1 pp 20--24 ; Physica D, 1986,
vol 18 pp 267--273,
 dedicated to the 60-th birthday of Martin Kruskal}


\bibitem{AC} Aharonov Y., Casher  A.
{\em  Phys Rev. A(3), 1979 vol 19 p 2461}

\bibitem{DN}Dubrovin B.A., Novikov S.P.
{\em Soviet Phys. JETP, 1980, vol 52.; Soviet Math Dokl. , 1980,
vol 253 p 1293}

\bibitem{SV}Shabat A.B., Veselov A.P.
{\em  Functional Analysis Appl., 1995 vol 29 iss 1}

\bibitem{W}Weiss J.
{\em Journ, Math Phys.,1987, vol 28(9) pp 2025--2039     }

\bibitem{DeV}de Vega H. J. , Schaposnik F.A.
{\em Phys Rev D14,  1100, 1976}

\bibitem{GMi}Gross D., Migdal A.A.
{\em  Princeton preprint PUPT--1159, 1989 }

\bibitem{BK}Brezin E., Kazakov V.
{\em   Preprint ENS, Paris,  1989  }

\bibitem{DS}Douglas M., Shenker S.
{\em   Rutgers preprint RU-89--34   }

\bibitem{D}David  F.
{\em    MPLA 5(13), 1019--1029, 1990 }

\bibitem{I}Its A.R. ,Novokshenov V.Yu.
{\em     Lecture Notes Math, vol 1191, Springer    }

\bibitem{FN}Flashka H., Newell A.
{\em     Comm Math Phys, 1980,  vol 76 p 67     }

\bibitem{Jap} Jimbo M., Miwa T., Ueno K.
{\em  Physica D vol 2 ,1981,  p 306 }

\bibitem{I2}Its A.R., Kitaev A.A.
{\em   MPLA, 1990, vol 5 iss 13 pp 1019--1029 }

\bibitem{Ko}Kapaev A.A.
{\em Differential Equations, 1988, vol 24    iss  10 p  1684 }

\bibitem{Ki}Kitaev A.V.
{\em Zap LOMI , 1991, vol  187  iss 12 p 53}


\bibitem{n2}Novikov S.P.
{\em Functional Analysis Appl.,1990, vol  24 iss  4 pp 43--53  }

\bibitem{Mo}Moore G.
{\em Comm Math Phys, 1990, vol 133  pp 261--304 }

\bibitem{Krich}Krichever I.M.
{\em  Preprint IHES, 1990}

\bibitem{IKF}Its A.R.,Fokas A.S., Kitaev A.V.
{\em   Comm Math Phys, 1992, vol 147 pp 395--430}


\bibitem{GN2}Grinevich P.G., Novikov S.P.
{\em Algebra and Analysis, 1994, in the volume, dedicated to the 60-th
birthday of L.D.Faddeev}

\end{thebibliography}
\end{document}